\documentclass[fleqn,10pt]{wlscirep}
\usepackage[utf8]{inputenc}
\usepackage[T1]{fontenc}
\usepackage{subfig}

\def\etal{\mbox{\it et al.\ }}

\def\ln{\textrm{ln}}
\newcommand{\eq}[1]{Eq.~\ref{#1}}
\newcommand{\fig}[1]{Fig.~\ref{#1}}
\newcommand{\tab}[1]{Table~\ref{#1}}

\title{A First-Principles-Based Approach to The High-Throughput Screening of Corrosion-Resistant High Entropy Alloys}

\author[1,*]{Thien Duong}
\author[3]{Yafei Wang}
\author[2]{Xiaoli Yan}
\author[3]{Adrien Couet}
\author[1, 2]{Santanu Chaudhuri}
\affil[1]{Applied Materials Division, Argonne National Laboratory, Lemont, Illinois, 60607}
\affil[2]{Department of Civil and Materials Engineering, University of Illinois - Chicago, Illinois, 60439}
\affil[3]{Department of Engineering Physics, University of Wisconsin - Madison, Wisconsin, 53706}
\affil[*]{tduong@anl.gov}

\begin{abstract}

    The design of corrosion-resistant high entropy alloys (CR-HEAs) is challenging due to the alloys' virtually astrological composition space. To facilitate this, efficient and reliable high-throughput exploratory approaches are needed. Toward this end, the current work reports a first-principles-based approach exploiting the correlations between work function, surface energy, and corrosion resistance (i.e., work function and surface energy are, by definitions, proportional and inversely proportional to an alloy's inherent corrosion resistance, respectively). Two Bayesian CALPHAD models (or databases) of work function and surface energy of \textit{fcc} Co-Cr-Fe-Mn-Mo-Ni are assessed using discrete surface energies and work functions derived by density-functional theory (DFT) calculations. The models are then used to rank different Co-Cr-Fe-Mn-Mo-Ni alloy compositions. It is observed that the ranked alloys possess chemical traits similar to previously studied corrosion-resistance alloys, suggesting that the proposed approach can be used to reliably screen HEAs with potentially good inherent corrosion resistance.

\end{abstract}

\begin{document}

\flushbottom
\maketitle

\thispagestyle{empty}


\section*{Introduction}

    As human civilization relies more and more on structural alloys, corrosion inevitably becomes a serious problem. Just within the US, the economic damage due to corrosion has materialized to 200-300 billion USD per year, or 3-4\% of the US gross domestic product (GDP) \cite{koch2002corrosion}. The figure has projected to 2-3 trillion USD per year worldwide, or 3-4\% global GDP, according to NACE International \cite{koch2016international}. Damages that have not been materialized yet include, for instance, the hindrance of advanced technologies such as the molten-salt technologies for energy generation and storage that possess high economic and environmental values. While common corrosion control practices can effectively reduce the impact of corrosion damage, the ultimate solution undoubtedly lies within the inventions of new alloys being superior to the conventional structural alloys.
    
    Recent advance in metallurgical research and development has resulted in a class of promising candidates, the high-entropy alloys (HEAs) \cite{tsai2014high, shi2017corrosion}. HEAs are alloys that have more than two principles components, are generally single-phase, and possess simple \textit{fcc}, \textit{bcc}, or \textit{hcp} structure. Their compositions were initially believed to be equiatomic or near-equiatomic such that their configurational entropies are highest possible and stabilize the alloys in simple single-phase structures. With later discoveries, however, the requirement of single-phase, equiatomic and/or near-equiatomic composition is not persisted and HEAs may also be referred as multi-principal element alloys (MPEAs) or complex concentrated alloys (CCAs) \cite{gorsse2017mapping}.
    
    Under the impression of the effect of configurational entropy, it is not surprising to see that the number of alloying elements could theoretically increase to an unconventional value and the number of equiatomic and near-equiatomic HEA compositions could consequently escalate to an extreme figure. With the relaxation of the equiatomic and near-equiatomic requirement, the number of possibly existing HEAs can only be astrological. With such a large school of alloys, shouldn't there proportionally be a myriad of unprecedented properties? In fact, it has been reported that many HEAs possess remarkable properties surpassing those of commercial alloys \cite{yeh2004nanostructured}; of these, many HEAs are excellent corrosion resistance alloys (CRAs) \cite{qiu2017corrosion}.
    
    Unfortunately, the alloys' advantages do not come without challenges. The large number of compositions makes it practically difficult to identify stable HEAs with desired properties. Conventional trial and error approach relying on experiments is cost prohibited. And as such, effective methods leveraging the advantages of modeling and simulation are needed.
    
    To reduce the material-design time and cost, the US Integrated Computational Materials Engineering (ICME) initiative \cite{national2008integrated} recommended the integration of multi-scale multi-physics modeling, experiment, and data science in 2008. Since then, it has seen various attempts to design materials following the ICME guidelines and successful stories have been gradually recorded. Of these, efforts to accelerate design of CR-HEAs using different ICME approaches are still scarce but have been being rigorously taken, without doubt. Lu Pin et. al. \cite{lu2018computational}, for instance, couples high-throughput (HT) CALPHAD analysis with the use of empirical pitting resistance equivalent number (a linearized fit of constituents' wt\%  to corrosion performance parameters) to narrow an extraordinary large number of HEA compositions down to only a few promising CR-HEAs. Among the predicted alloys, one has been verified by experiment. It should be noted that Lu Pin et. al. \cite{lu2018computational} focused specifically on single-phase HEAs (v.s. multiple-phase HEAs) due presumably to the following reasons: (1) it simplifies design criterion therefore enhance screening accuracy, and (2) single-phase alloys are less susceptible to Galvanic corrosion.
    
    The current work contributes a similar HT-screening approach relying, however, on \textit{ab initio} achievable indicators instead of empirical indicators. Namely, surface energy and work function. We also target single-phase HEAs with focus on the \textit{fcc} phase, for the same aforementioned reasons. Our hypothesis is that surface energy and work function are highly correlated to the corrosion resistance of an alloy and as such can be used to rank the alloy's corrosion-resistance potent in relative to other alloy compositions.
    
    To further interpret, surface energy indicates how strong/stable a surface of a material is and as such can be used to infer how susceptible the material is against a thermodynamic reaction/chemical attack. A material with a low surface energy could, in this regard, be implied to have a better corrosion resistance than a material with a high surface energy. Work function represents how difficult it is to move an electron from the surface of a material to vacuum above the surface. To some extent, it is related to Galvanic corrosion and also represents the tendency towards the formation of ionic/covalent bonds between surface atoms of a materials and surrounding environmental substances (i.e. the very core of the redox reaction). A material with a higher work function, as such, could be implied to have a stronger corrosion resistance than a material with a lower work function.
    
    The surface energies and work functions of Co-Cr-Fe-Mn-Mo-Ni alloys are assessed using HT-DFT calculations. To make it systematic, the surface energies and work functions of low-order \textit{fcc} alloys (up to quaternary) with different alloy compositions belonging to the senary Co-Cr-Fe-Mn-Mo-Ni system are calculated. We then propose the use of the CALPHAD methodology \cite{saunders1998calphad, lukas2007computational}, specifically the sub-regular solution model, to model the populated first-principles data. The CALPHAD models allow smooth interpolations across the alloy composition domain as well as extrapolations to higher-order \textit{fcc} HEAs, allowing to achieve more information for less data. In addition, as no model is inert to error, Bayesian statistics is adopted to quantify, to some extent, the uncertainties of the CALPHAD models \cite{2016Duong}.
    
    As a study case, the assessed Bayesian CALPHAD models are used to rank the inherent corrosion-resistance potent of various equiatomic and near-equiatomic \textit{fcc} HEAs belonging to the Co-Cr-Fe-Mn-Mo-Ni system. It is found that equiatomic and near-equiatomic alloys featuring high-Ni, and high-Co and/or high-Mo tend to exhibit good inherent corrosion resistance while those with high-Cr and/or high-Mn are predicted to be more susceptible to corrosion attacks. These findings are in good phenomenological agreement with previous experimental studies of commercial alloys and in consistent with the current Inductively Coupled Plasma Mass Spectrometry (ICP-MS) experiments taken to further verify the approach. They suggest that the proposed approach can be exploited to facilitate the design of CR-HEAs.

\section*{Results}

    \subsection*{Bayesian CALPHAD surface energy and work function databases}
    
    Extensive Bayesian CALPHAD databases of surface energy and work function of Co-Cr-Fe-Mn-Mo-Ni are assessed using various DFT-derived surface energies and work functions of discrete alloy compositions. The details of the DFT calculations and the Bayesian CALPHAD modeling are reported in the Method section. It is noted here that, due to the high computational cost of the DFT calculations, the raw surface energies and work functions are derived mostly for mechanically stable unaries, except those being available from Tran et. al. \cite{tran2016surface}, and binaries. For the interested Cr-Fe-Mn-Mo-Ni sub-system, surface energies and work functions are additionally populated for a limited number of ternary and quaternary alloys. The whole range of alloy composition is accounted, with a composition step of 25 at.\% being considered for binary and ternary alloys and a composition step of 12.5 at.\% being considered for quaternary alloy. Of the generated data, it should be noted that a few surface energies and work functions belonging to mechanically unstable fcc alloys are excluded. This is due to the fact that their values are relatively far off from those of neighboring alloys, raising the likelihood that the DFT calculations are not well converged (or being trapped locally).
    
\begin{figure}[!ht]
    	\begin{center}
    	    \includegraphics[width=0.50\columnwidth]{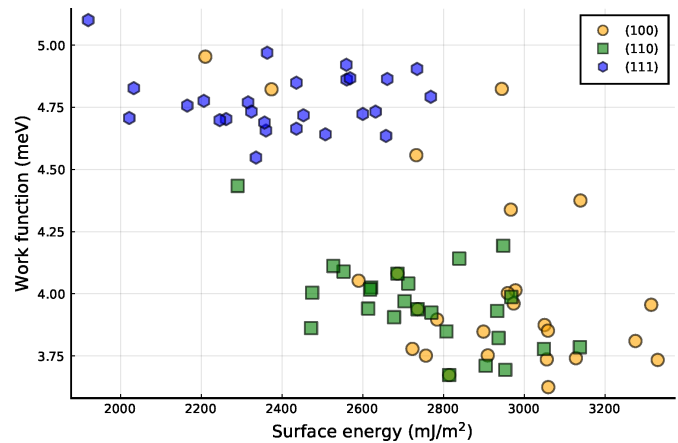}
    		\caption[CrFeMnNi Choice of surface]{Calculated (111), (110), and (100) surface energies and work functions of different Cr-Fe-Mn-Ni alloys indicate that the (111) surface tend to possess the best combination of surface energy and work function.}
    		\label{fig:CrFeMnNi_hkl_choice}
    	\end{center}
    \end{figure}

    
    In addition, we focus specifically on the (111) surface since it possesses the best combination of surface energy and work function among the three common (100), (110), and (111) indices of the \textit{fcc} structure. To further interpret this, \fig{fig:CrFeMnNi_hkl_choice} shows various DFT-derived (100), (110), and (111) surface energies and work functions of different Cr-Fe-Mn-Ni alloy compositions. As can be seen from this figure, (111) generally has the lowest surface energy and highest work function among the studied surfaces. Since work function and surface energy of a material are proportional and inversely proportional to the material's inherent corrosion resistance respectively, the predicted values indicate that the (111) surface tends to possess the best corrosion resistance among the three common orientations of a \textit{fcc} alloy

    The Bayesian CALPHAD models (or databases) are implemented via a user-friendly web interface that is available at https://cradle.egs.anl.gov/. The interface and databases are actively maintained and developed. Currently, surface energies and work functions of additional alloy compositions are being populated for improved accuracy. Additional alloying elements as well as properties relevant to the design of CR-HEAs will be added in the future. The following discusses the application of the CALPHAD models in the HT-screening of CR-HEAs belonging to the Co-Cr-Fe-Mn-Mo-Ni system. It should be noted that these results and discussions are based on the current version of the CALPHAD databases. With future developments, minor changes are possible.

    \subsection*{Screening of CR-HEAs}
    
    \tab{tab:CoCrFeMnMoNi_Best10} and \tab{tab:CoCrFeMnMoNi_Worst10} report our predictions of the 10 best and the 10 worst Co-Cr-Fe-Mn-Mo-Ni compositions in terms of their inherent corrosion resistances, respectively. Taking into accounts the modeling uncertainties, the ranking of these alloys is conducted in the following manner. Firstly, 100,000 plausible parameter sets of each CALPHAD model are sampled using the MCMC approach (see the method section for more details). For each parameter set, the surface energies or work functions of various alloy compositions across the senary composition domain are evaluated. To avoid the case in which the resulting best-10 and worst-10 groups would be crowded with alloys that possess similar compositions, the composition step is chosen to be 0.2 at$\%$ for each alloying element. The alloy compositions are ranked based on the distances between their estimated surface energies/work functions and predefined references. The references are chosen to be 1000 $mJ/m^2$ and 6000 $meV$ for surface energy and work function, respectively. The alloy that has the smallest distances from the chosen surface energy and work function is deemed to possess the highest inherent corrosion resistance while the alloy that has the farthest distances from the chosen surface energy and work function is considered the most susceptible to corrosion attacks. The 10 best and worst alloy compositions are recorded for each parameter set and the ranking is repeated for the total of 100,000 sampled parameter sets. This results in 1,000,000 best recorded compositions and 1,000,000 worst recorded compositions. The histograms of these two sets of recorded compositions point to the alloy compositions that possess the highest likelihood of being among the best 10 and the worst 10 compositions, according to the Bayesian statistics.
    
    As can be seen from \tab{tab:CoCrFeMnMoNi_Best10} and \tab{tab:CoCrFeMnMoNi_Worst10}, the alloys that are predicted to possess potentially good inherent corrosion resistance are mostly those with high-Co and/or high-Ni content while the alloys that are predicted to be potentially bad at inherent corrosion resistance are mostly those possessing high-Mn and/or high-Cr content. Alloys that have high Mo content, as it will be shown later, are also predicted to be less susceptible to corrosion attacks while alloys with high Fe content can be either potentially good or bad CRAs depending on the relative amounts of Co, Cr, Mn, and/or Ni that they possess.
    
    \begin{table}[htbp]
        \begin{center}
            \caption{Best 10 ranked compositions belonging to the Co-Cr-Fe-Mn-Mo-Ni system. Note that the ranking is based on the likelihood that the alloy belongs to the top 10 compositions with the best combination of surface energy and work function; it does not necessarily mean that the 1st ranked composition has a better combination of surface energy and work function than the last ranked composition in the list.}
            \begin{tabular}{|r|r|r|r|r|r|}
                \hline
                \multicolumn{1}{|l|}{Co (at \%)} & \multicolumn{1}{l|}{Cr (at \%)} & \multicolumn{1}{l|}{Fe (at \%)} & \multicolumn{1}{l|}{Mn (at \%)} & \multicolumn{1}{l|}{Mo (at \%)} & \multicolumn{1}{l|}{Ni (at \%)} \\ \hline
                    0 & 0 & 0.2 & 0 & 0 & 0.8 \\ \hline
                    0.8 & 0 & 0.2 & 0 & 0 & 0 \\ \hline
                    0 & 0 & 0.4 & 0 & 0 & 0.6 \\ \hline
                    1 & 0 & 0 & 0 & 0 & 0 \\ \hline
                    0 & 0 & 0 & 0 & 0 & 1 \\ \hline
                    0.6 & 0 & 0.4 & 0 & 0 & 0 \\ \hline
                    0 & 0 & 0.6 & 0 & 0 & 0.4 \\ \hline
                    0 & 0.2 & 0 & 0 & 0 & 0.8 \\ \hline
                    0 & 0 & 0 & 0 & 0.2 & 0.8 \\ \hline
                    0 & 0 & 0.2 & 0.2 & 0 & 0.6 \\ \hline
            \end{tabular}
            \label{tab:CoCrFeMnMoNi_Best10}
        \end{center}
    \end{table}

    \begin{table}[htbp]
        \begin{center}
            \caption{Worst 10 ranked compositions belonging to the Co-Cr-Fe-Mn-Mo-Ni system. Note that the ranking is based on the likelihood that the alloy belongs to the bottom 10 compositions with the worst combination of surface energy and work function; it does not necessarily mean that the 1st ranked composition has a worse combination of surface energy and work function than the last ranked composition in the list.}
            \begin{tabular}{|r|r|r|r|r|r|}
                \hline
                \multicolumn{1}{|l|}{Co (at \%)} & \multicolumn{1}{l|}{Cr (at \%)} & \multicolumn{1}{l|}{Fe (at \%)} & \multicolumn{1}{l|}{Mn (at \%)} & \multicolumn{1}{l|}{Mo (at \%)} & \multicolumn{1}{l|}{Ni (at \%)} \\ \hline
                    0 & 1 & 0 & 0 & 0 & 0 \\ \hline
                    0 & 0.8 & 0.2 & 0 & 0 & 0 \\ \hline
                    0.2 & 0.8 & 0 & 0 & 0 & 0 \\ \hline
                    0 & 0.8 & 0 & 0.2 & 0 & 0 \\ \hline
                    0 & 0 & 0.6 & 0.4 & 0 & 0 \\ \hline
                    0 & 0 & 0.4 & 0.6 & 0 & 0 \\ \hline
                    0 & 0.8 & 0 & 0 & 0.2 & 0 \\ \hline
                    0 & 0.6 & 0.4 & 0 & 0 & 0 \\ \hline
                    0.2 & 0.6 & 0 & 0 & 0.2 & 0 \\ \hline
                    0 & 0.6 & 0 & 0.4 & 0 & 0 \\ \hline
            \end{tabular}
            \label{tab:CoCrFeMnMoNi_Worst10}
        \end{center}
    \end{table}
    
    \begin{figure}[!ht]
    	\begin{center}
    	    \subfloat[a]{\includegraphics[width=0.45\columnwidth]{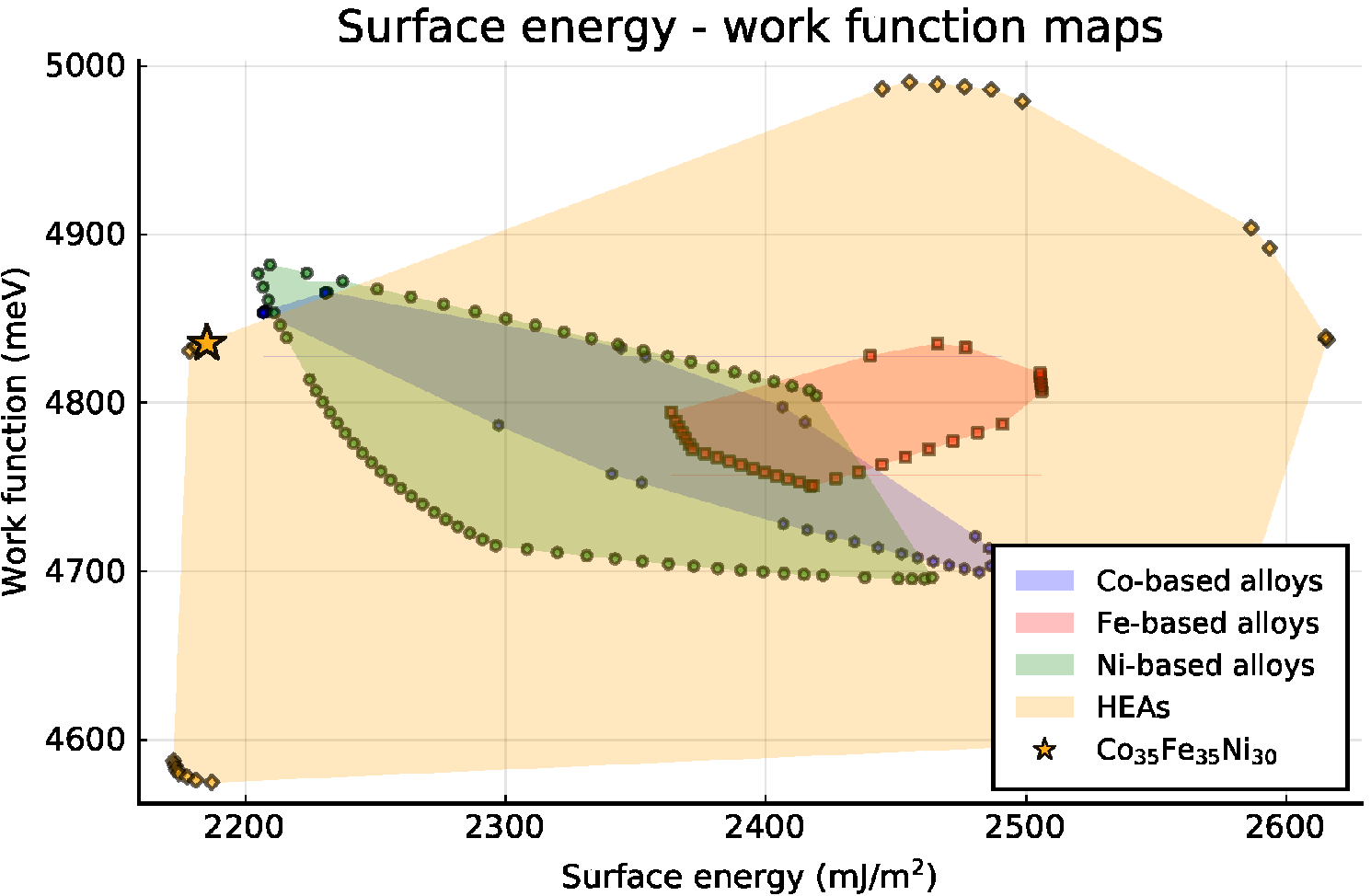}}
    	    \subfloat[b]{\includegraphics[width=0.45\columnwidth]{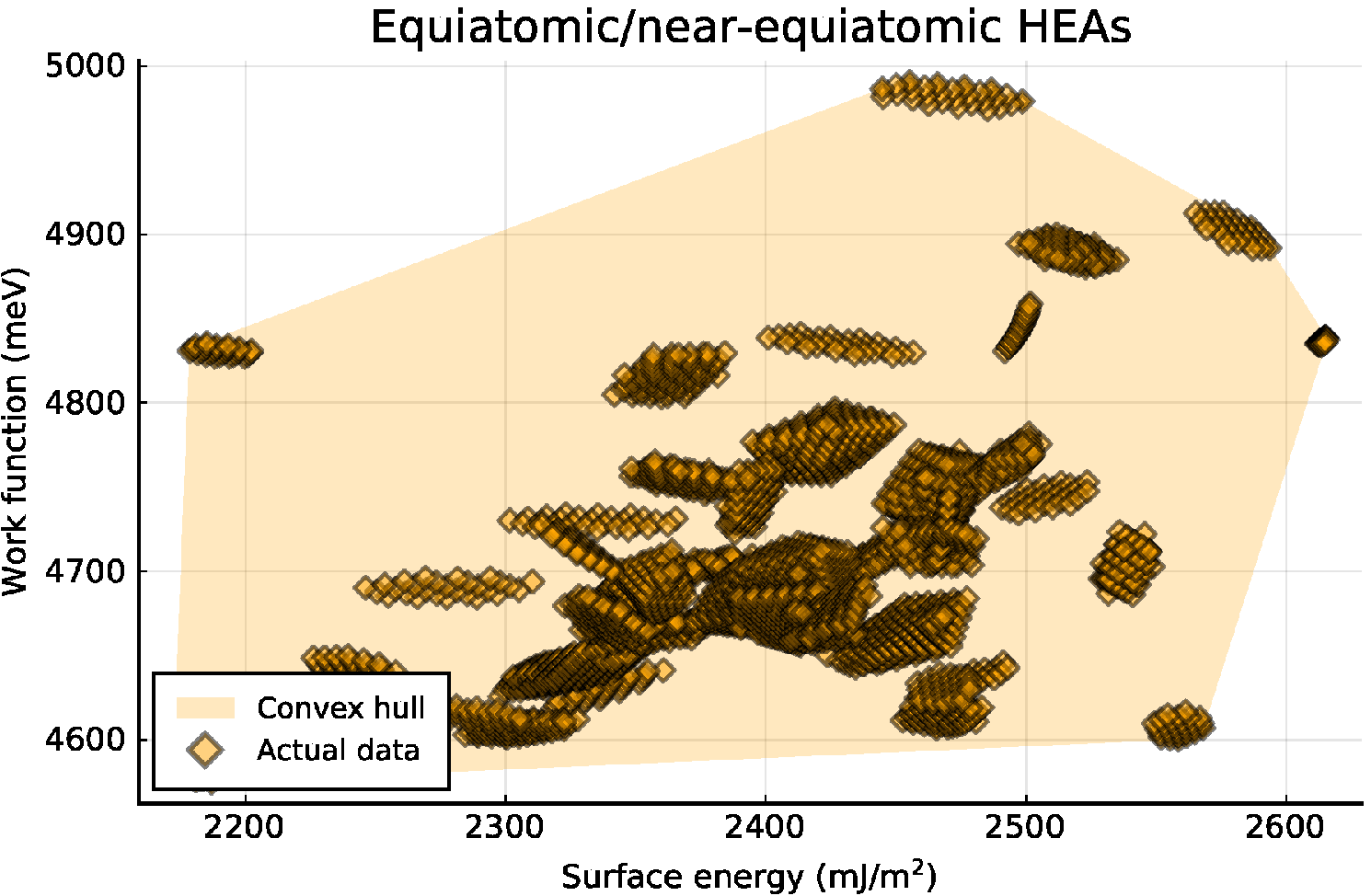}}
    		\caption[CoCrFeMnMoNi surface-work]{(a) The convex-hull maps of surface energies and work functions of simulated equiatomic, near-equiatomic HEAs, and conventional alloys and (b) Estimated surface energies and work functions of equiatomic and near-equiatomic HEAs. Simulated equiatomic and near-equiatomic HEAs are alloys with at least 3 alloying elements and with compositions being within 5 at\% deviation from each other. Simulated Fe-based and Co-based alloys are alloys with compositions approximating those of stainless steels and Co-based superalloys, respectively. Simulated Ni-based alloys are alloys with compositions approximate those of Ni-based superalloys, albeit without Al and W.}
    		\label{fig:CoCrFeMnMoNi_alloy_map}
    	\end{center}
    \end{figure}
    
    \fig{fig:CoCrFeMnMoNi_alloy_map} shows the 2-D convex-hull maps of surface energies and work functions of simulated equiatomic, near-equiatomic HEAs, and some conventional alloys. It is noted that, due to limited visualization capability, only the mean values of the Bayesian predictions are captured and reported. Additionally, the convex hull is adopted for a better visualization experience. Due to the nature of the simulated HEA compositions, concealed points within the HEA hull are not uniformly distributed across the hull but rather in scattered clusters (see \fig{fig:CoCrFeMnMoNi_alloy_map} (b) for demonstration). The simulated equiatomic, near-equiatomic HEAs are alloys with at least 3 alloying elements and with elemental concentrations being no more than 5 at\% different from each other. The simulated conventional alloys are (a) Co-based alloys, which simulate Co-based superalloys, (b) Fe-based alloys, which simulate stainless steels, and (c) Ni-based alloys, which resemble Ni-based superalloys, albeit without the common Al and W elements. The compositions of the simulated conventional alloys are reported in \tab{tab:simulated_alloys}.
    
    \begin{table}[htbp]
        \caption{Simulated Co-based alloys, Fe-based alloys (steels), and Ni-based alloys}
        \begin{center}
            \begin{tabular}{|c|c|c|c|c|c|c|}
                \hline
                 & Co (at \%) & Cr (at \%) & Fe (at \%) & Mn (at \%) & Mo (at \%) & Ni (at \%) \\ \hline
                Co-based alloys & balance & 0.2 – 0.3 & 0 – 0.05 & 0 – 0.02 & 0 – 0.1 & 0 – 0.1 \\ \hline
                Fe-based alloys & 0 – 0.01 & 0.2 – 0.3 & balance & 0 – 0.05 & 0 – 0.01 & 0.1 – 0.2 \\ \hline
                Ni-based alloys & 0.1 – 0.3 & 0.1 – 0.3 & 0 – 0.01 & 0 – 0.01 & 0 – 0.05 & balance \\ \hline
            \end{tabular}
        \end{center}
        \label{tab:simulated_alloys}
    \end{table}

    ``\textit{Why HEAs?}'' The common believe is that the large combinatorial number of alloy compositions infers a wide range of properties within which superior properties exist. This has been supported by the gradually increasing number of studies of existing and/or new HEAs. There are, however, scarce quantitative demonstrations of how large the property space of HEAs could be in proportion to their composition space. \fig{fig:CoCrFeMnMoNi_alloy_map} provides such an illustration. Here, it can be seen from the figure that, in relative to the simulated HEAs, the property spaces of the simulated conventional alloys are somewhat restricted. This is due to the fact that they are biased by the properties of the principle elements. HEAs, on the other hand, exhibit a relatively broader property space, despite the fact that their composition domain, like those of the simulated conventional alloys, are also restricted (i.e. vicinity of equiatomic composition v.s. vicinity of principle components' compositions). Apparently, the more complicated interactions among the alloying elements in HEAs result in more complex yet more widespread properties.
    
    Based on the predicted surface energies and work functions of the simulated alloys, we identify that the Co-Fe-Ni HEAs possess the most balanced surface energy and work function combination, among the simulated alloys. Their surface energies and work functions occupy a small area around the marked Co$_{35}$Fe$_{35}$Ni$_{30}$ alloy in \fig{fig:CoCrFeMnMoNi_alloy_map}. Assuming that properties other than corrosion resistance of an alloy are also desired, the trade-off would likely be the alloy's lower corrosion resistance potent. Within the context of this work, this is translated to a higher surface energy and a lower work function (than those specified by the Co-Fe-Ni alloys). By specifying the surface energy and work function thresholds to be 2300 ($mJ/m^2$) and 4600 ($meV$) respectively, equiatomic and near equiatomic Fe-Mo-Ni and Fe-Mn-Mo-Ni alloys are additionally identified. These predicted alloys indicate that Mo, similar to Co and Ni, is also a favored alloying element for CRAs. 
    
\section*{Discussion}

    \subsection*{Rationale of the proposed approach}
    
    The correlations between work function, surface energy, and corrosion resistance of solid solutions have been well studied. Marattukalam et. al. \cite{marattukalam2015microstructure}, for instance, investigated the correlation between the surface energy and the corrosion resistance of NiTi shape-memory alloy. The alloy was fabricated using Laser Engineered Net Shaping (LENS) and its corrosion behaviors and surface energies as functions of LENS's processing parameters were measured. It was found by the authors that a decrease in surface energy shifted measured corrosion potential to nobler direction and decreased measured corrosion current. Jinlong et. al. \cite{jinlong2016effect} studied the effect of grain orientation on the corrosion resistance of nanocrystalline pure nickel. The authors found that the (111) orientation plan of pure nickel increased significantly the metal's corrosion resistance, regardless of the metal's grain size. It was argued that the highest corrosion resistance of the (111) surface was associated with the surface's lowest energy among the common (100), (110), and (111) orientations of the \textit{fcc} metal. Schmutz et. al. \cite{schmutz1998characterization} measured the work functions of pure metals in air and found that they were proportional to the open-circuit potentials in aqueous solutions. Wang et. al. \cite{wang2013high} attributed the enhanced high-temperature oxidation resistance of BN-SS305 stainless steel in both initial and iso-thermal oxidation to a larger work function and more chemical stability. Shi et. al. \cite{shi2018homogenization} studied the effect of homogenization on the corrosion behaviors of the multi-phase Al$_x$CoCrFeNi HEAs and concluded that the homogenization resulted in microstructure simplification which in turn decreased the work function variations among existing phases and resulted in the alloys' improved corrosion resistance.
    
    These studies support the fact that work function and surface energy are proportional and inversely proportional to corrosion resistance of materials -- the correlations that can be inferred from surface energy's and work function's definitions. The same correlations are observed in the current work, for the specific case of Co-Cr-Fe-Mn-Mo-Ni alloys.
    
    Indeed, the above CR-HEA rankings based on predicted surface energy and work function suggest that alloys with high Co, Mo, and/or Ni contents generally possess high inherent corrosion resistance while alloys with high Cr and/or Mn contents generally possess low inherent corrosion resistance. These are indeed the facts that have been established by previous experiments. For instance, Olson et. al. \cite{olson2009materials} measured the corrosion of various alloys, mostly industrial alloys, at 850$^{0}C$ for 500 hours. Among the studied alloys, a majority contains Co, Cr, Fe, Mn, Mo, and/or Ni, albeit the amounts of Mn are trivial. The authors observed that alloys with high initial Cr tended to be more susceptible to corrosion while high-Ni alloys generally possessed lower corrosion rates. Examples were Ni-201 with highest Ni content (99.4 wt.$\%$Ni) and lowest weight lost and Haynes-230 with highest Cr content (22.5 wt.$\%$Cr) and highest weight lost. Compared to Haynes-230, Inconel-617 had a relatively similar composition but with higher Co, higher Mo, less Ni, and no W. The authors considered Mo and W as corrosion-resistance enhancing elements, based on previous findings \cite{misra1987proceedings}. It was observed that Inconel-617 experienced a considerably lower weight loss than that of Haynes-230. This suggested that Inconel-617's higher Co and Mo contents had well compensated for the potential lost of the alloy's corrosion resistance caused by its lower Ni content and the lack of W. The authors believed that the corrosion behaviours of the studied alloys were thermodynamic-driven. To support their argument, the authors cited the calculated formation enthalpies (per mole fluorine) of the alloys' prospective corrosion products. The thermodynamic quantities implied that Cr is the most susceptible to corrosion attacks while Ni, Co, and Mo are the better alloying elements to combat corrosion attacks. In a recent study \cite{elbakhshwan2019corrosion}, the corrosion of near-equiatomic Cr$_{18}$Mn$_{27}$Fe$_{27.5}$Ni$_{27.5}$ alloy in molten FLiBe was studied at 700$^{0}C$ for 1000 hours. Unlike the previously studied alloys, Cr$_{18}$Mn$_{27}$Fe$_{27.5}$Ni$_{27.5}$ is compositionally concentrated with a notably high Mn content. The authors observed that the near-equiatomic alloy had a higher mass lost compared to the reference 316H stainless steel. This was due mostly to the dissolution of Mn into the molten salt. It was believed that this dissolution of Mn had discouraged the dissolution of Cr which was commonly observed in previous studies\cite{olson2009materials}. The prolonged thermal exposure of the alloys at 700$^{0}C$ had resulted in decomposition into a \textit{bcc} + \textit{fcc} matrix, which could in turn lead to more complex corrosion mechanisms. However, it could be argued that the early state of the alloy's corrosion was thermodynamic-driven, similar to Olson et. al.\cite{olson2009materials}.
    
    
    These studies are in line with the current predictions based on the estimated surface energies and work functions of Co-Cr-Fe-Mn-Mo-Ni alloys. Here, it is noted that even though formation enthalpies can be used to phenomenologically explain alloys' corrosion behaviors, these information alone is insufficient to map alloys' compositions to alloys' corrosion for the HT-screening purpose. The use of surface energy and work function, on the other hand, allows direct unambiguous ranking of alloys' inherent corrosion-resistance potent. This indicates that surface energy and work function are higher-level data that contain richer and/or more useful information.
    
    \subsection*{Experimental verification}
    
\begin{figure}[!ht]
    	\begin{center}
    	    \subfloat[]{\includegraphics[width=0.45\columnwidth]{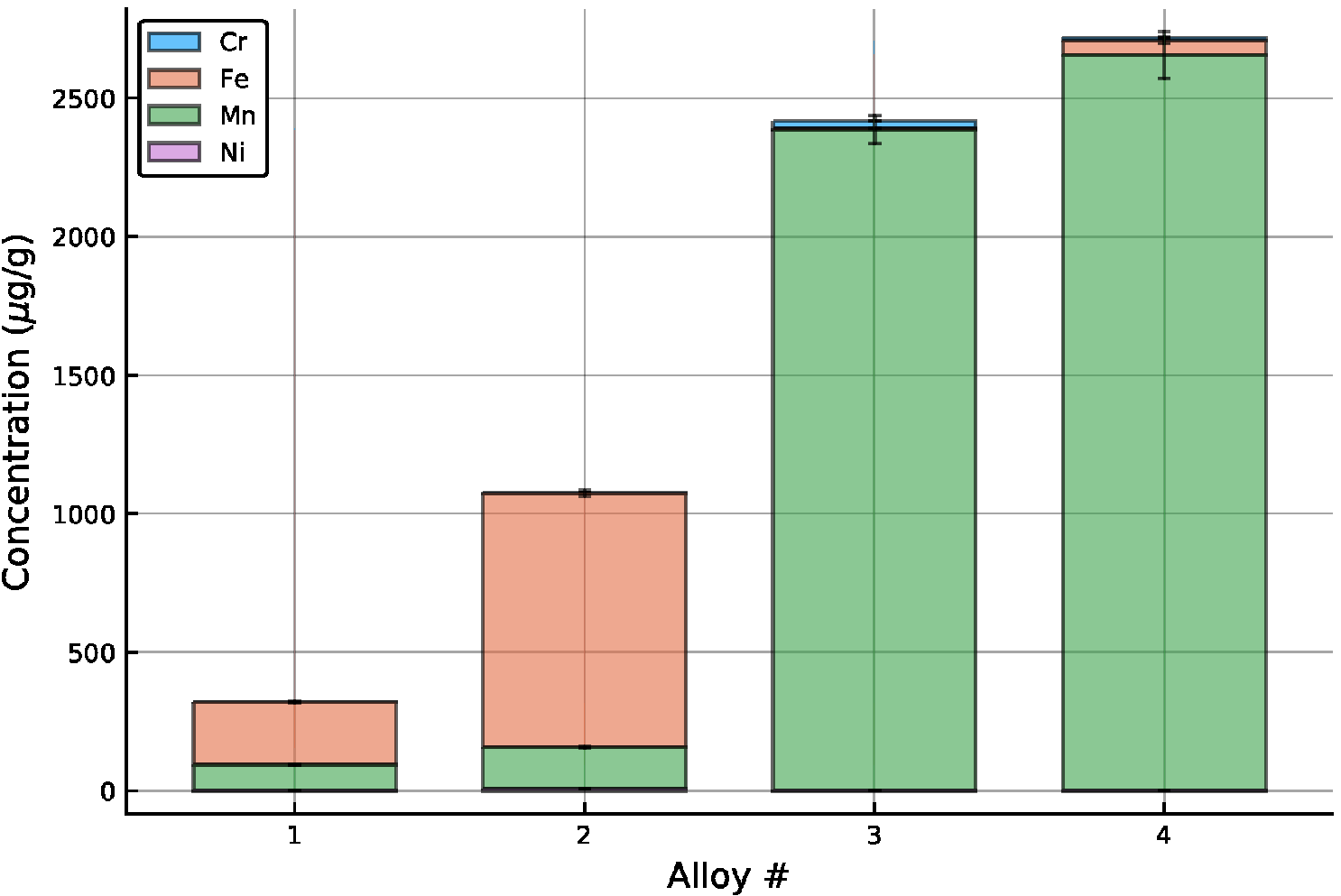}}
    	    \subfloat[]{\includegraphics[width=0.45\columnwidth]{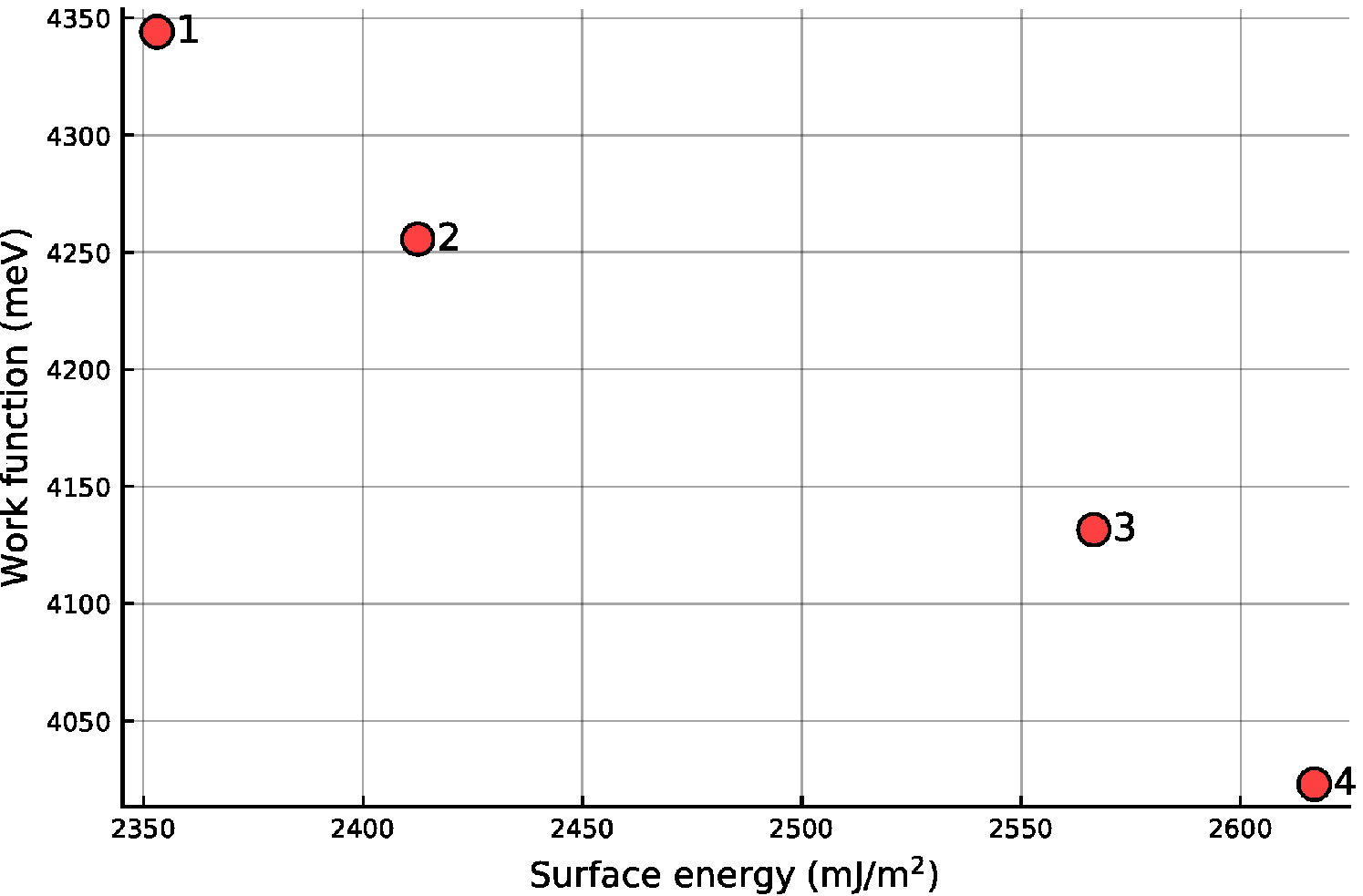}}
    		\caption[exp. vs. comp.]{(a) Concentrations of the dissolved ions in the molten salt for each of the tested alloy after corrosion experiment. (b) Calculated mean surface energy and work function based on the compositions of the 4 tested alloys (110 plane).}
    		\label{fig:vs_experiment}
    	\end{center}
    \end{figure}

    To further verify the proposed HT-screening approach, corrosion experiments were carried out to identify the corrosion resistances of four Ni-rich alloys from the Cr-Fe-Mn-Ni alloy system. The compositions of the alloys were measured by energy-dispersive X-ray spectroscopy (EDS) and listed in \tab{tab:EDS_composition}. The alloys were additively manufactured using in-situ alloying in a direct energy deposition system, followed by homogenization at 1000°C and aging at 700°C. The details of alloy processing can be found in \cite{Phalgun2021} All four alloys were confirmed to be single FCC phase by X-Ray Diffraction. The sample texture was mostly (110) plane paralleling to the sample surface \cite{Yafei2021}. Samples were mirror polished for the corrosion experiment. The corrosion experiment was performed at 500 $^0C$ for 96 hours in LiCl-KCl eutectic molten salt with 2 wt.\% of EuCl$_3$ added to increase the salt's corrosiveness. It was conducted inside a glovebox filled with inert argon atmosphere to avoid/reduce exposure to external impurities. During the corrosion of alloys in molten salt environment, the unstable elements in the alloy dissolved into the molten salt in the forms of ions and formed metal chlorides \cite{guo2018corrosion}. By measuring the concentrations of the dissolved ions in the post-corrosion salt using ICP-MS, the corrosion resistances of different alloys could be quantified. In turn, the corrosion-resistance ranking of the four tested alloys was established. The obtained ICP-MS results for each post corrosion salt are shown in \fig{fig:vs_experiment} (a). The trace impurities of the pre-corrosion salt were also identified by ICP-MS analysis in which the ten most prominent impurity elements and their concentrations were 16.5 ppm Na,7.5 ppm Mg, 1.1 ppm Al, 3.8 ppm P, 11.2 ppm S, 32.3 ppm Ca, 0.2 ppm Cr, 1.0 ppm Fe, 0.02 ppm Mn, and 0.1 ppm Ni. As can be seen from \fig{fig:vs_experiment} (a), the alloy \#1 has the lowest dissolved ions content in the salt after 96 hours, followed by alloy \#2, alloy \#3, and alloy \#4. This means alloy \#1 is the most corrosion resistant alloy while alloy \#4 is the worst among the 4 tested alloys. As can be seen from \fig{fig:vs_experiment} (b), this result is in remarkable agreement with the ranking order established based on the predicted (110) surface energies and work functions. Here, the (110) surface energies and work functions are predicted using Bayesian CALPHAD models that are developed using the aforementioned DFT data and method. The excellent agreement between the experimental results and the proposed approach lend us the believe that the proposed theoretical approach is reliable for HT-screening of CR-HEAs. 
    
    \begin{table}[htbp]
        \caption{Compositions of the alloys used in the corrosion experiments}
        \begin{center}
            \begin{tabular}{|r|r|r|r|r|}
                \hline
                \multicolumn{ 1}{|l|}{Sample number} & \multicolumn{ 4}{c|}{Composition (at.\%)} \\ \cline{ 2- 5}
                \multicolumn{ 1}{|l|}{} & \multicolumn{1}{l|}{Cr} & \multicolumn{1}{l|}{Fe} & \multicolumn{1}{l|}{Mn} & \multicolumn{1}{l|}{Ni} \\ \hline
                1 & 0.6 & 10.4 & 1.1 & 87.9 \\ \hline
                2 & 1.2 & 18.4 & 2.0 & 78.4 \\ \hline
                3 & 1.3 & 12.6 & 19.3 & 66.7 \\ \hline
                4 & 0.9 & 30.5 & 14.7 & 53.9 \\ \hline
            \end{tabular}
        \end{center}
        \label{tab:EDS_composition}
    \end{table}

    \subsection*{Factors affecting screening accuracy}
    The accuracy of the HT-screening depends on how reliable predicted surface energies and work functions are. This depends, in succession, on the reliability of CALPHAD extrapolations and how these predictions fare against ground truth under the impacts of temperature and/or environment. In the following, we assess the accuracy of CALPHAD extrapolation, the impacts of temperature and environment on surface energy and work function, and how they may affect the resulted alloy ranking based on the proposed approach.

    Regarding CALPHAD extrapolations, since the increase in the calculated DFT data is not necessarily proportional to the increase in the complexity of the higher-order alloy system, CALPHAD extrapolations' accuracy is of concern. To assess the reliability of CALPHAD extrapolations, additional DFT calculations are conducted to evaluate the surface energies and work functions of randomly chosen Co$_{12.5}$Cr$_{12.5}$Fe$_{12.5}$Mn$_{50}$Mo$_{12.5}$, Co$_{12.5}$Cr$_{12.5}$Fe$_{12.5}$Mn$_{12.5}$Ni$_{50}$, and Co$_{62.5}$Cr$_{12.5}$Fe$_{12.5}$Mn$_{12.5}$. Like in the previous DFT calculations, the additional alloys' bulk structures are modeled using 32-atom special-quasi structures (SQSs) and their surface structures are modeled using 96-atom SQSs. The results of the calculations are shown in \tab{tab:extrapolation} in comparisons with Bayesian CALPHAD predictions. Here, for each alloy, the predicted min-max range and 95$\%$ confidence interval (CI), shown in the table, are derived from predictions that use 100,000 plausible parameter sets sampled by the MCMC approach (see the Method section for details).
    
    
    \begin{table}[htbp]
        \caption{DFT surface energy and work function v.s. CALPHAD-predictions}
        \begin{center}
            \begin{tabular}{|l|l|r|l|l|}
                \hline
                Alloys & Property & DFT & CALPHAD Min-Max & CALPHAD 95$\%$ CI \\ \hline
                \multicolumn{ 1}{|l|}{Co$_{12.5}$Cr$_{12.5}$Fe$_{12.5}$Mn$_{50}$Mo$_{12.5}$} & Surface energy (mJ/m$^2$) & 2460 & 2230 -- 2655 & 2433 -- 2570 \\ \cline{ 2-5}
                \multicolumn{ 1}{|l|}{} & Work function (meV) & 4950 & 4222 -- 5105 & 4674 -- 4907\\ \hline
                \multicolumn{ 1}{|l|}{Co$_{12.5}$Cr$_{12.5}$Fe$_{12.5}$Mn$_{12.5}$Ni$_{50}$} & Surface energy (mJ/m$^2$) & 2163 & 2029 -- 2502 & 2216 -- 2334 \\ \cline{ 2-5}
                \multicolumn{ 1}{|l|}{} & Work function (meV) & 5197 & 4251 -- 5175 & 4670 -- 4820\\ \hline
                \multicolumn{ 1}{|l|}{Co$_{62.5}$Cr$_{12.5}$Fe$_{12.5}$Mn$_{12.5}$} & Surface energy (mJ/m$^2$) & 2149 & 2097 -- 2455 & 2204 -- 2345 \\ \cline{ 2-5}
                \multicolumn{ 1}{|l|}{} & Work function (meV) & 4890 & 4356	-- 5392 & 4825 -- 5055\\ \hline
            \end{tabular}
        \end{center}
        \label{tab:extrapolation}
    \end{table}
    
    As can be seen from \tab{tab:extrapolation}, the CALPHAD-predicted Min-Max ranges are mostly in good agreement with the DFT calculations. With the narrower 95$\%$ CI ranges, while the predicted surface energies are still in good agreement with the DFT calculations, most of the DFT-derived work functions are outside of the ranges. The deviations from the predicted bounds are, however, reasonably small except for the case of Co$_{12.5}$Cr$_{12.5}$Fe$_{12.5}$Mn$_{12.5}$Ni$_{50}$ with a deviation of $\sim8\%$ from the predicted bound.
    
    Regarding the impact of temperature, it is noted that the current surface energies and work functions are evaluated at ground-state condition. At elevated temperatures, Gibbs free energy would change due to the effect of entropy. Since Gibbs free energy is correlated to surface energy and work function, the absolute values of surface energy and work function would also change. This, however, would not necessarily result in the change of the order of ranked alloys. Indeed, among the energetic components of Gibbs free energy, enthalpy is usually the much larger one at low and intermediate temperatures (see for example \cite{schon2018probing} in which the role of enthalpy in HEAs is highlighted). It follows that the differences in Gibbs free energies among the ranked alloys at low and intermediate temperatures would be decided by the enthalpy differences. Since the contribution of enthalpy in the total Gibbs free energy is much less dependent on temperature (than entropy), the relativity among the ranked alloys' enthalpies are likely established at the ground-state (0 K) condition and remain throughout the low to intermediate temperatures. Unless for a very high temperature, entropy changes due to temperature increase are usually not large enough to disrupt the order established by the enthalpy differences at low and intermediate temperatures.
    
\begin{figure}[!ht]
    	\begin{center}
    	    \includegraphics[width=0.5\columnwidth]{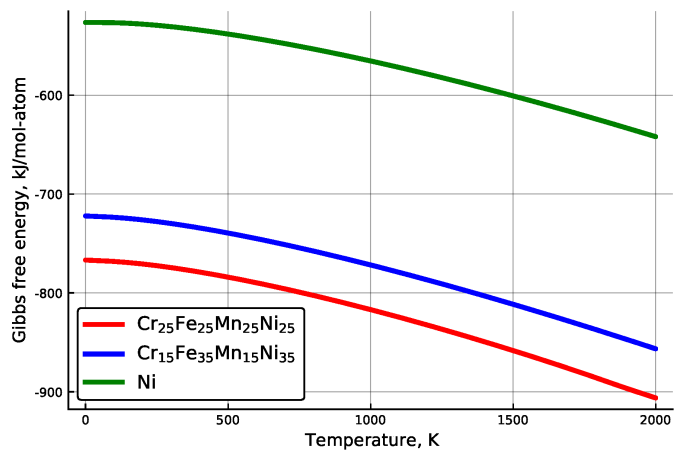}
    		\caption[CoCrFeMnMoNi surface-work]{DFT-derived Gibbs free energies of \textit{fcc} Ni, equiatomic CrFeMnNi, and Cr$_{15}$Fe$_{35}$Mn$_{15}$Ni$_{35}$ bulks as functions of temperature. The plot indicates that entropy differences among different alloy compositions do not alter the thermodynamic relativity initially defined by the alloys' enthalpy differences (at 0 K) even at temperature as high as 2000 K.} 
    		\label{fig:T_dependent_energy}
    	\end{center}
    \end{figure}

    To demonstrate, \fig{fig:T_dependent_energy} shows the T-independent thermodynamic relativity of different bulk compositions. Here, the T-depedent Gibbs free energies of \textit{fcc} Ni, equiatomic CrFeMnNi, and Cr$_{15}$Fe$_{35}$Mn$_{15}$Ni$_{35}$ bulks are calculated using the supercell approach \cite{Wei1992} within the quasi-harmonic approximation considering both vibrational and electronic contributions. For the cases of the disordered alloys, 4-atom and 20-atom special quasi-random structure (SQS) \cite{Zunger1990} are used to model the alloys, respectively. The ATAT package \cite{Vandewalle2017} is used to generate the SQSs and to setup VASP calculation. The details of VASP calculations can be found in the Method section. Configurational entropies of the alloys are assumed ideal and hence described by the classical $RT \sum_{i} x_{i}\ln(x_{i})$ formula. As can be seen from \fig{fig:T_dependent_energy}, the thermodynamic relativity among the three bulk alloys remain almost unchanged even though there exists differences among the alloys' entropies (including electronic, vibration, and configuration). This is due to the fact that the magnitudes of the entropy differences, even though are magnified by temperature, are relatively much smaller than the magnitudes of the enthalpy differences among the compositions, which are mainly defined at 0K.
    
    Although the thermodynamic properties of bulk and surface structures are different, they follow the same fundamentals reflected by the Gibbs' thermodynamic relations. And as such, we expect that the role of enthalpy difference in surface structures is similar to that in bulk structures. To reiterate, since the Gibbs free energy of a surface structure is correlated to the structure's surface energy and work function, it is believed that the relative order among the ground-state surface energies and work functions of different alloy compositions generally remains at low and intermediate temperatures.
    
    \begin{table}[htbp]
    \caption{Calculated surface energies and work functions of some Cr-Fe-Mn-Mo-Ni alloys in vacuum and water salt environments}
        \begin{center}
            \begin{tabular}{|l|l|c|c|}
                \hline
                \multicolumn{ 1}{|l|}{Composition} & \multicolumn{ 1}{l|}{Property} & \multicolumn{ 2}{c|}{Environment} \\ \cline{ 3- 4}
                \multicolumn{ 1}{|l|}{} & \multicolumn{ 1}{l|}{} & Vacuum & Salt Water \\ \hline
                \multicolumn{ 1}{|l|}{Cr$_{0.5}$Fe$_{0.5}$} & Surface Energy (mJ/m$^2$) & 2660 & 2600 \\ \cline{ 2- 4}
                \multicolumn{ 1}{|l|}{} & Work Function (meV) & 4860 & 4610 \\ \hline
                \multicolumn{ 1}{|l|}{Cr$_{0.25}$Fe$_{0.25}$Ni$_{0.5}$} & Surface Energy (mJ/m$^2$) & 2206 & 2161 \\ \cline{ 2- 4}
                \multicolumn{ 1}{|l|}{} & Work Function (meV) & 4780 & 4530 \\ \hline
                \multicolumn{ 1}{|l|}{Cr$_{0.125}$Fe$_{0.125}$Mn$_{0.125}$Ni$_{0.625}$} & Surface Energy (mJ/m$^2$) & 2165 & 2152 \\ \cline{ 2- 4}
                \multicolumn{ 1}{|l|}{} & Work Function (meV) & 4760 & 4470 \\ \hline
                \multicolumn{ 1}{|l|}{Cr$_{0.125}$Fe$_{0.625}$Mn$_{0.125}$Ni$_{0.125}$} & Surface Energy (mJ/m$^2$) & 2436 & 2400 \\ \cline{ 2- 4}
                \multicolumn{ 1}{|l|}{} & Work Function (meV) & 4850 & 4610 \\ \hline
                \multicolumn{ 1}{|l|}{Cr$_{0.25}$Fe$_{0.5}$Mo$_{0.25}$} & Surface Energy (mJ/m$^2$) & 2464 & 2427 \\ \cline{ 2- 4}
                \multicolumn{ 1}{|l|}{} & Work Function (meV) & 4750 & 4460 \\ \hline
            \end{tabular}
        \end{center}
    \label{tab:salt_water_surfwork}
    \end{table}
    
    Lastly, \tab{tab:salt_water_surfwork} compares the surface energies and work functions of some randomly selected alloys that are calculated in salt water environment using the implicit solution model \cite{mathew2014implicit} with those calculated in vacuum. For efficiency, the cavitation energy contribution in the implicit solution model is neglected \cite{mathew2019implicit}. It can be seen from \tab{tab:salt_water_surfwork} that surface energies and work functions are generally smaller in the salt water environment than in vacuum. This is due to the former's higher relative permittivity. The relative ranking among the alloy compositions, however, remains unchanged for the studied cases.
    
    The above studies show that the accuracy of CALPHAD extrapolation and the effects of temperature and environment may impact the accuracy of the proposed HT-screening approach but expectedly not by a large degree. Given that the nature purpose of HT-screening is to narrow a large solution space down to a practical size (rather than to find the very best solution), we believe that the potential impacts are tolerable and the proposed screening approach is a good pre-processing for the design process of CR-HEAs. Alternatively, the assessed surface energy and work function databases can be exploited in more sophisticated machine-learning approaches that utilize various sources of information (or features) to better infer promising CR-HEAs.
    
    
\section*{Methods}

\subsection*{Density-functional theory}

    The Vienna Atomistic Simulation Package (VASP) \cite{Kresse1996} was used to perform \textit{ab initio} calculations. Studied alloys were modeled using the special quasi-random structure (SQS) \cite{Zunger1990} implemented in the ATAT package \cite{Vandewalle2017}. Specifically, 32-atom SQSs were used to model \textit{fcc} bulk alloys and 96-atom supercells, generated from the SQSs using the pymatgen package \cite{tran2016surface}, were used to model surface structures. The calculations were performed with Generalized Gradient Approximation (PBE) projector augmented wave pseudopotentials \cite{Perdew1996}. More specifically, VASP pv versions of the potentials were considered when available. Collinear spin polarization was accounted. Integration in the reciprocal space were obtained over a $\tau$-centered Monkhorst-Pack grid in which the number of $k$ points was chosen to be $N_{k point}\approx3000/N_{atom}$ within the first Brillouin zone \cite{Monkhorst1976}. Cutoff energy was set to 1.3 $\times$ highest constitutional ENMAX. The electronic and ionic convergence criteria were set to $10^{-6}/10^{-5}eV$ respectively for bulk calculations. For surface calculations, lower accuracy criteria of $10^{-5}/10^{-3}$ were chosen respectively for electronic and ionic convergence for a lower cost. A few test cases, using the higher accuracy criteria of $10^{-6}/10^{-5}eV$, show that the choice of lower accuracy criteria does result in discrepancies but the differences are reasonably small. Relaxations were realized using the Hermite-Gauss smearing method of Methfessel and Paxton of order 1, with a smearing parameter of 0.01 eV\cite{Methfessel1989}. The bulk structures were fully relaxed while only atomic degree of freedom were considered in the relaxations of surface structures. For improved accuracy, final static calculations using the  the tetrahedron smearing method with Blöchl corrections \cite{blochl1994improved} were considered for all calculations.
    
    In addition, the Exact Muffin-Tin Orbital (EMTO) approach \cite{vitos2007computational} was adopted to evaluate the elastic constants and, in turn, the mechanical stability of the studied Co-Cr-Fe-Mn-Mo-Ni alloys. The DFT-derived surface energies and work functions of alloys that are identified as being mechanical stable are weighted 10 times higher than those of alloys that are identified as mechanical unstable in the Bayesian CALPHAD assessment. The computational details of EMTO calculations can be found in \cite{2016Duong}.
    
\subsection*{CALPHAD modeling}

    To interpolate the missing surface energies and work functions of uncovered compositions, we adopt the CALPHAD methodology. Specifically, the subregular solution model is used to model the hypothesized corrosion indicators:
    
    \begin{equation}
        P = \sum_{i}x_{i}{P^{0}_{i}} + \sum_{i,j}x_{i}x_{j}P_{i,j} + \sum_{i,j,k}x_{i}x_{j}x_{k}P_{i,j,k} 
        \label{eq:GibbsFreeEnergy}
    \end{equation}
    where, the temperature-dependent parametric terms are omitted from the solution model due to use of ground-state DFT data, $P$ is either surface energy or work function, $x_{i}$ is the mole fraction, $P^{0}_{i}$ is the surface energy or work function of pure elements, $P_{i,j}$ represents contributions beyond the Vegard's law and is represented in term of Redlich-Kister polynomial: $P_{i,j} = \sum\limits_{\nu=0}^{n}{P^{i,j}_{\nu}}(x_{Nb}-x_{U})^{\nu}$, where $n$ is currently chosen to be 0, $P^{i,j}_{\nu}$ and $P_{i,j,k}$ are parametric constants.
    
    The parameters were estimated by solving \eq{eq:GibbsFreeEnergy}, given enough known $P$ values. The error varies for different properties with the highest max error and MSE being $\sim15\%$ and $\sim3\%$ of the original DFT data, respectively. To better account for the model uncertainty, Bayesian uncertainty quantification is adopted.
    
\subsection*{Bayesian Uncertainty Quantification and Propagation}

    The Matlab MCMC toolbox by Haario \etal \cite{haario2001adaptive, haario2006dram} was adopted to quantify the uncertainties of the model parameters, within the Bayesian statistics. The priors and likelihoods of model parameters were assumed to be uniform and Gaussian distributions, respectively. Following previous work \cite{2016Duong}, we set, as empirical priori, the ranges of the uniform distributions to be from $-3 \times Par$ to $+3 \times Par$, where $Par$ are the original parameters estimated by solving \eq{eq:GibbsFreeEnergy}. The variances of the Gaussian distributions were treated as hyper-parameters and were initially set to 0.05. The MCMC simulation sampled 100,000 promising parameter candidates according to the Metropolis-Hastings algorithm. During the MCMC sampling process, the variances of the Gaussian likelihoods were dynamically updated to better describe the likelihood distributions. The collected parameter sets were used to derived the means and standard deviations of predicted surface energies and work functions.

\bibliography{sample}

\section*{Acknowledgements}

The calculations were conducted using Argonne National Laboratory's LCRC and AFCL clusters. The authors thank the ARPA-E for its support under contract number PRJ1007310.

\section*{Author contributions}

T.D. and S.C. conceived the hypothesis-based screening approach, T.D. conducted the DFT calculations and Bayesian CALPHAD modeling and analysed the results. Y.W. and A.C. conducted ICP-MS experiments and analysed the results. All authors reviewed the manuscript. 

\section*{Competing interest}

\section*{Additional information}

\textbf{Correspondence} and requests for materials should be addressed to T.D.




\end{document}